\documentclass{sig-alternate}

\usepackage[utf8]{inputenc}

\usepackage{amsfonts}

\usepackage{listings}
\usepackage{float}
\usepackage{color}
\usepackage{colortbl}
\usepackage{footnote}
\makesavenoteenv{tablefootnote}

\definecolor{KWColor}{rgb}{0.37,0.08,0.25}
\definecolor{CommentColor}{rgb}{0.133,0.545,0.133}
\definecolor{StringColor}{rgb}{0,0.126,0.941}

\usepackage{tikz} 
\usetikzlibrary{decorations.text}
\usetikzlibrary{patterns, automata}
\usetikzlibrary{calc,trees,positioning,arrows,chains,shapes.geometric,%
    decorations.pathreplacing,decorations.pathmorphing,shapes,%
    matrix,shapes.symbols}
\usetikzlibrary{arrows}
\usepackage{multirow}    
\usepackage{hhline}      

\usepackage{graphicx}
\usepackage{caption}
\usepackage{subcaption}
\usepackage{soul}
\usepackage{algorithm} 
\usepackage{algpseudocode}

\usepackage{datapie}

\newcommand*\hit{%
  \scalebox{.65}{
  \begin{tikzpicture}
    \node[draw,circle,inner sep=.9pt,text width=5pt] {\scalebox{1.15}{$\star$}};
  \end{tikzpicture}}
  }

\newcommand*\fp{%
  \scalebox{.65}{
  \begin{tikzpicture}
    \node[inner sep=.9pt,text width=5pt,double,color=red] {\scalebox{1.4}{$\star$}};
  \end{tikzpicture}}
  }

\newcommand*\fn{%
  \scalebox{.8}{
  \begin{tikzpicture}
    \node[draw,circle,inner sep=.9pt,text width=5pt,color=red] {};
  \end{tikzpicture}}
  }

\usepackage{array}
\usepackage{booktabs}
\usepackage{rotating}

\usepackage{balance}

\newcolumntype{P}[2]{%
  >{\begin{turn}{#1}\begin{minipage}{#2}\small\raggedright\hspace{0pt}}c%
  <{\end{minipage}\end{turn}}%
}

\usepackage{pgfplotstable}
\usepackage{siunitx}
\usetikzlibrary{patterns}

\input{pie-macros}
\makeatletter
\long\def\pgfplots@addplotimpl@coordinates@#1#2#3#4{%
    \pgfplots@start@plot@with@behavioroptions{#1,/pgfplots/.cd,#2}%
    \pgfplots@PREPARE@COORD@STREAM{#4}%
    \begingroup
    \edef\@tempa{{#3}}%
    \ifpgfplots@curplot@threedim
        \expandafter\endgroup\expandafter
        \pgfplots@coord@stream@foreach@threedim\@tempa
    \else
        \expandafter\endgroup\expandafter
        \pgfplots@coord@stream@foreach\@tempa
    \fi
}%
\makeatother

\newcommand{\chartFromFile}[3][]{

\pgfplotstableread[col sep=comma, ignore chars={"}]{#3}{\barctable}
\pgfplotstablegetrowsof{#3}
\pgfmathsetmacro{\rows}{\pgfplotsretval-1}  
\pgfplotstablegetcolsof{#3}
\pgfmathsetmacro{\columns}{\pgfplotsretval-1}

\gdef\yNames{}
\foreach \i in {0,...,\rows}{%
  \pgfplotstablegetelem{\i}{[index] 0}\of{\barctable} 
  \let\cName\pgfplotsretval 
  \xdef\yNames{\yNames\ifx\yNames\empty\else,\fi\cName}
}

\pgfplotsset{
/pgfplots/xbar legend/.style={
/pgfplots/legend image code/.code={%
\draw[##1,/tikz/.cd,bar width=3pt,yshift=-0.2em,bar shift=0pt]
plot coordinates {(0cm,0.8em) };},
},
/pgfplots/ybar legend/.style={
/pgfplots/legend image code/.code={%
\draw[##1,/tikz/.cd,bar width=3pt,yshift=-0.2em,bar shift=0pt]
plot coordinates {(0cm,0.8em) };},
}
}

\begin{tikzpicture}
\footnotesize

 \begin{axis}[
  nodes near coords={\tiny\pgfmathprintnumber\pgfplotspointmeta\%},
  yticklabel={\num[round-mode=places, round-precision=0]{\tick}\%},
  bar width=10pt,
  height=5cm,
  width=.95\columnwidth,
  ymin=0,
  ymax=100,
  ybar=2*\pgflinewidth,
  xtick=data,
  symbolic x coords/.expand once={\yNames},
  ymajorgrids = true,
  x tick label style={text width=1.4cm, align=center},
  legend style = {cells={anchor=west}},
  ]

\foreach \j in {1,...,\columns} {
  
  \global\def\mycoordinates{}

  \foreach \i in {0,...,\rows}{%
    \pgfplotstablegetelem{\i}{[index] 0}\of{\barctable} 
    \let\cName\pgfplotsretval 
    \pgfplotstablegetelem{\i}{[index] \j}\of{\barctable} 
    \let\cj\pgfplotsretval 

    \xdef\mycoordinates{\mycoordinates (\cName,\cj)}

  }

  \addplot coordinates \mycoordinates;
  
  \pgfplotstablegetcolumnnamebyindex{\j}\of{\barctable}\to{\colname}
  \addlegendentryexpanded[align = left]{\colname}
}

\end{axis}
\end{tikzpicture}

} 

\newcommand{\tool}[1]{IccTA}

\usepackage{amssymb,amsmath}
\newboolean{showcomments}
\setboolean{showcomments}{true}
\ifthenelse{\boolean{showcomments}}
{ \newcommand{\mynote}[2]{
    \fbox{\bfseries\sffamily\scriptsize#1}
    {\small$\blacktriangleright$\textsf{\emph{#2}}$\blacktriangleleft$}}}
{ \newcommand{\mynote}[2]{}}

\newcommand{\precisionDroidbench}[1]{95.0\%}
\newcommand{\recallDroidbench}[1]{82.6\%}
\newcommand{\numberOfDroidbenchApps}[1]{26}
\newcommand{\topICCMethods}{8}

\begin{document}

\title{I know what leaked in your pocket: uncovering privacy leaks on Android Apps with  Static Taint Analysis}

\author{
{\rm Li Li, Alexandre Bartel, }\\ 
\rm Jacques Klein, Yves Le Traon\\
\rm SnT\\
\rm University of Luxembourg\\
\rm {firstName.lastName@uni.lu}
\and
{\rm Steven Arzt, Siegfried Rasthofer,}\\ 
\rm and Eric Bodden\\
\rm EC SPRIDE\\ 
\rm Technische Universit\"at Darmstadt\\ 
\rm {firstName.lastName@ec-spride.de}
\and
{\rm Damien Octeau, Patrick McDaniel}\\
\rm Department of Computer Science and\\ 
\rm Engineering\\ 
\rm Pennsylvania State University\\
\rm \{octeau,mcdaniel\}@cse.psu.edu
}

\maketitle

\begin{abstract}

Android applications may leak privacy data carelessly or maliciously.
In this work we perform inter-component data-flow analysis to detect privacy leaks between components of Android applications.
Unlike all current approaches, our tool, called IccTA, propagates the context between the components, which improves the precision of the analysis.
IccTA outperforms all other available tools by reaching a precision of \precisionDroidbench{} and a recall of \recallDroidbench{} on DroidBench.
Our approach detects 147 inter-component based privacy leaks in 14 applications in a set of 3000 real-world applications with a precision of 88.4\%.
With the help of ApkCombiner, our approach is able to detect inter-app based privacy leaks.

\end{abstract}

 \section{Introduction}
\label{introduction}

With the growing popularity of Android, thousands of applications (also called apps) emerge every day on the official Android market (Google Play) as well as on some alternative markets.
As of May 2013, 48 billion apps have been installed from the Google Play store, 
and as of September 3, 2013, 1 billion Android devices have been activated~\cite{androidwiki}.
Researchers have shown that Android apps frequently send the user's private data outside the device without the user's prior consent~\cite{zhou2012dissecting}.
Those applications are said to \emph{leak} private data.
Android applications are made of different components; most of the privacy leaks are simple and operate within a single component.
More recently, cross-component and also cross-app privacy leaks have been reported~\cite{wu2013impact}.
Analyzing components separately is not enough to detect such leaks.
Therefore, it is necessary to perform an inter-component analysis of applications.
Android app analysts could leverage such a tool to identify malicious apps that leak private data.
For the tool to be useful, it has to be highly precise and minimize the false positive rate when reporting applications leaking private data.

\textbf{Privacy leaks.}
In this paper, we use a static taint analysis technique to find privacy leaks, i.e., paths from sensitive data, called \emph{source}s, to statements sending the data outside the application or device, called \emph{sink}s.
A path may be within a single component or cross multiple components and/or applications.

State-of-the-art approaches using static analysis to detect privacy leaks on Android apps mainly focus on detecting intra-component sensitive data leaks. 
CHEX~\cite{chex}, for example, uses static analysis to detect component hijacking vulnerabilities by tracking taints between sensitive sources and sinks.
DroidChecker~\cite{chan:droidchecker} uses inter-procedural Control-Flow Graph (CFG) searching and static taint checking to detect exploitable data paths in an Android application.
FlowDroid~\cite{steven2014pldi} also performs taint analysis within single components of Android applications but with a better precision.
In this paper,
we not only focus on intra-component leaks, but we also consider Inter-Component Communication (ICC) based privacy leaks, including Inter-Application Communication (IAC) leaks.

Other approaches use dynamic tracking to find privacy leaks.
For instance, TaintDroid~\cite{enck2010taintdroid} leverages Android's virtualized execution environment to monitor Android apps at runtime in which it tracks how application leaks private information.
CopperDroid~\cite{reina2013system} dynamically observes interactions between the Android components and the underlying Linux system to reconstruct higher-level behavior.

A dynamic approach must send input data to the app at runtime to trigger code execution.
The input data may be incomplete and thus not execute all parts of the code.
Furthermore, some code may only be executed if precise conditions are met at runtime such as a data.
In this paper, we focus on static analysis to avoid these drawbacks.
The counterpart of static analysis is that it may yield an over-approximation since it analyzes all code even the one that could never be executed.

\textbf{Static taint analysis for Android is difficult.} 
Despite the fact that Android applications are mainly programmed in Java, off-the-shelf static taint analysis tools for Java do not work on Android applications.
The tools need to be adapted mainly for three reasons. 
The first reason is that, as already mentioned, Android applications are made of components. 
Communications between components involve two main artifacts:  \textit{Intent Filter} and \textit{Intent}.  
An \textit{Intent Filter} is attached to a component and ``filters'' \textit{Intents} that can reach the component.
An \textit{Intent} is used to start a new component by first dynamically creating an \textit{Intent} instance, and then by calling a specific method (e.g. \textit{startActivity}, \textit{startService}) with the \textit{intent} previously created as parameter. The \textit{intent} is used either explicitly by specifying the new component to call, or implicitly by for instance only specifying the action\footnote{Such as \texttt{android.intent.action.VIEW} or \texttt{.CALL} or \texttt{.EDIT}} to perform.    
The launch of a component is performed by the Android system which ``resolves'' the matching between \textit{Intent} and \textit{Intent Filter} at runtime. 
This dynamic resolution done by the Android system induces a discontinuity in the control-flow of Android applications. 
This specificity makes static taint analysis challenging by requiring pre-processing of the code to resolve links between components.

The second reason is related to the user-centric nature of Android applications, in which a user can interact a lot through the touch screen. 
The management of user inputs is mainly done by handling specific callback methods such as the \textit{onClick} method which is called when the user clicks on a button. 
Static analysis requires a precise model that stimulates users' behavior.
 
The third and last reason is related to  the lifecycle management of the components. 
There is no \textit{main} method as in a traditional java program. 
Instead, the Android system switches between states of a component's lifecycle by calling callback methods such as \textit{onStart}, \textit{onResume} or \textit{onCreate}.
However, these lifecycle methods are not directly connected in the code. 
Modeling the Android system allows to connect callback methods to the rest of the code.

\textbf{Our Proposal.}
The above challenges will unavoidably cause some discontinuities in the control-flow graph. 
To overcome these issues, 
we present an Inter-component communication Taint Analysis tool named IccTA\footnote{Our experimental results and IccTA itself are available at https://sites.google.com/site/icctawebpage.}.
IccTA allows a sound and precise detection of ICC and IAC links.
This approach is generic and can be used for any data-flow analysis.
In this paper we focus on using IccTA to detect privacy leaks.

IccTA is based on three software artifacts: Epicc-IccTA, FlowDroid-IccTA and ApkCombiner.

Epicc-IccTA extends Epicc~\cite{octeau2013effective} which computes ICC links between Android components.
Epicc-IccTA leverages Epicc to incrementally store the computed ICC links to a database for conveniently analyzing a large set of apps.
FlowDroid-IccTA extends FlowDroid~\cite{steven2014pldi}.
FlowDroid only finds privacy leaks within single components of Android applications but not between components.

FlowDroid-IccTA uses ICC links computed by Epicc to improve FlowDroid.
Based on these computed links, Flow\-Droid-IccTA modifies Android applications' code to directly connect components to enable data-flow analysis between components.
By doing this, we build a complete control-flow graph of the whole Android application.
This allows propagating the context between Android components and yielding a highly precise data-flow analysis.
To the best of our knowledge, this is the first approach that precisely connects components for data-flow analysis.

Finally, ApkCombiner helps analyzing multiple Android applications by combining multiple apps into one when there exist data flows between these apps.
This results in having a complete control-flow graph of the combined apps.
This allows to propagate the context not only between components of a single app but also between components of different apps.

To verify our approach, we developed \numberOfDroidbenchApps{} apps containing ICC based privacy leaks.
We have added these applications to DroidBench~\cite{droidbench}, an open test suite for evaluating the effectiveness and accuracy of taint analysis tools specifically for Android apps.
The \numberOfDroidbenchApps{} apps cover the top \topICCMethods{} used ICC methods illustrated in Table~\ref{tab:icc_methods}.

\textbf{Contributions.}
To summarize, we present the following original contributions in this paper:
\begin{itemize}

\item A novel methodology to resolve the ICC problem by directly connecting the discontinuities of Android apps at the code level.

\item IccTA, a tool for inter-component data-flow analysis.

\item An improved version of DroidBench with \numberOfDroidbenchApps{} new apps to evaluate tools detecting ICC based privacy leaks.

\item An empirical study to evaluate IccTA over an augmented version of the DroidBench test suite (available online\footnote{github.com/secure-software-engineering/DroidBench}) and 3000 real-world Android applications.

\end{itemize}
 
 \section{Background}
\label{background}

\subsection{Android ICC Methods}

An Android application is made of basic units, called components, described in a special file, called \emph{Manifest}, stored in the application.
There are four types of components: 
a) Activities that represent the user interface and are the visible part of Android applications; 
b) Services which execute tasks in background; 
c) Broadcast Receivers that receive messages from other components or the system, such as incoming calls or text messages; 
and d) Content Providers which act as the standard interface to share structured data between applications. 

Some specific Android system methods are used to trigger inter-component communication. 
We call them Inter-Component Communication (ICC) methods. 
Those methods take as parameter a special kind of object, called \emph{Intent}, which specifies the target component(s).
We perform a short study to compute the usage rate of ICC methods. 
We analyzed 3000 Android applications randomly selected from Google Play and other third party markets.
Table~\ref{tab:icc_methods} shows the top \topICCMethods{} most used ICC methods.
The third column represents the number of apps using at least once the corresponding ICC method.
The most used ICC method is \texttt{startActivity}, used to launch a new Activity component, which accounts for 59.2\% of the total detected ICC methods.

\begin{table}
\definecolor{LightGray}{gray}{0.9}
\caption{The top \topICCMethods{} used ICC methods$^{\dagger}$}
\label{tab:icc_methods}
\footnotesize
\begin{tabular}{ p{7.1pc}  p{5.2pc}  p{5.1pc} }
\hline
ICC Method             			& Counts(\#.)   & Used Apps(\#.) \\  \hline
\rowcolor{LightGray} 
startActivity        			& 55802 ($61.44\%$)  	& 2765 ($92.2\%$) \\ 
startActivityForResult			& 11095 ($12.21\%$)  	& 1980 ($66.0\%$) \\ 
\rowcolor{LightGray} 
query					& 6606  ($7.27\%$)	& 1601 ($53.4\%$) \\
startService          			& 3942  ($4.34\%$) 	& 1077 ($35.9\%$) \\
\rowcolor{LightGray} 
sendBroadcast         			& 3472  ($3.82\%$)  	& 790 ($26.3\%$) \\ 
insert					& 2100  ($2.31\%$)	& 615 ($20.5\%$) \\
\rowcolor{LightGray}
bindService  				& 1515  ($1.67\%$)  	& 644 ($21.5\%$) \\  
delete					& 1238   ($1.36\%$)	& 350 ($11.7\%$) \\
\rowcolor{LightGray} 	
Other ICC Methods 			& 5058  ($5.57\%$)	& -			\\
Total					& 90828 ($100\%$) 	& - 			 \\ 
\hline
\end{tabular}
{\footnotesize
\parbox{\columnwidth}{$^{\dagger}$ Methods with higher counts are selected when overload methods exist} \\
}
\end{table}

All ICC methods\footnote{Except \texttt{Content Provider} related methods such as \texttt{query} or \texttt{insert}} take at least one \textit{Intent} in their parameters to specify the target component(s). 
There are two ways to specify ICC method's target components.
The first one is by explicitly specifying them by setting the name of the target components through an \textit{Intent}.
The second one is by implicitly specifying them by setting the \emph{action}, \emph{category} and \emph{data} fields of an \textit{Intent}.
In order to receive implicit \textit{Intents}, target components need to specify an \textit{Intent Filter} in their application's manifest file.
Note that \textit{Intents} can transfer data between components.

Again, we performed a short study on the 3000 apps to compute the ratio between explicit and implicit \textit{Intents} for the \texttt{startActivity} ICC method.
Among the 55,802 \texttt{startActivity} method calls, 27978 use explicit \textit{intents} and 27824 use implicit \textit{Intents}.

\begin{figure}
\begin{center}
\resizebox{\columnwidth}{!}{
 \begin{tikzpicture}
\tikzstyle{act} = [ellipse, draw, inner sep = 1mm, minimum width=2cm];
\tikzstyle{app} = [draw, dotted, rounded corners];
  \path 
  (0,0) node[act] (a1a1) {Activity$_1$}
++(2,-1) node[act] (a1a2) {Activity$_2$}
    +(-1,0) node[app, minimum width=4.5cm, minimum height=3.5cm, label={[label distance=-.45cm]90:\footnotesize Application 1}] {}
++(-2,-1) node[act, label={[label distance=-.1cm]-90:\color{red}{\small IF: Action$_b$}}] (a1a3) {Activity$_3$}
++(5, 2) node[act, label={[label distance=-.1cm]-90:\color{red}{\small IF: Action$_b$}}] (a2a1) {Activity$_4$} 
    +(0,0) node[app, minimum width=2.5cm, minimum height=1.5cm, label={[label distance=-.45cm]90:\footnotesize Application 2}] {}
++(0,-2) node[act, label={[label distance=-.1cm]-90:\color{red}{\small IF: Action$_c$}}] (a3a1) {Activity$_5$} 
    +(0,0) node[app, minimum width=2.5cm, minimum height=1.5cm, label={[label distance=-.45cm]90:\footnotesize Application 3}] {}
++(-6,-1.4) node (la) {}
++(1,0) node (lb) {}
++(1,0) node[text width=2cm] (l) {\footnotesize Explicit ICC \\ to Activity$_2$}
++(2,0) node (ra) {}
++(1,0) node (rb) {}
++(1,0) node[text width=2cm] (r) {\footnotesize Implicit ICC \\ for action$_b$}
;

\draw[-latex] (a1a1) -- (a1a2);
\draw[-latex, dashed] (a1a2) to[out=0, in=-170, distance=1.5cm] (a2a1);
\draw[-latex, dashed] (a1a2) to[out=0, in=-20, distance=1.5cm] (a1a3);
\draw[-latex] (la) -- (lb);
\draw[-latex, dashed] (ra) -- (rb);
 \end{tikzpicture}}
\end{center}
\caption{Explicit and Implicit ICC between Components of Android Applications.}
\label{fig:icc_graph}
\end{figure}
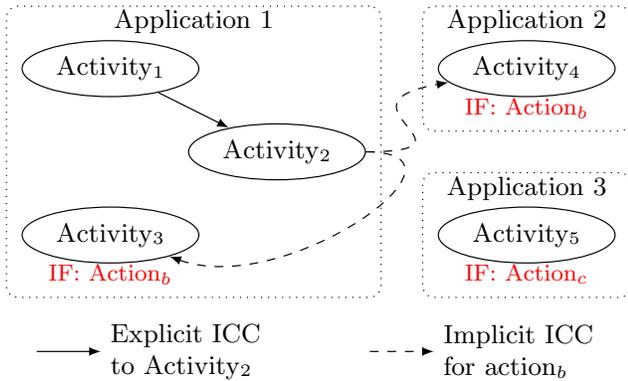

Figure~\ref{fig:icc_graph} represents three Android apps made of Activity components.
There is an explicit ICC from \texttt{Activity$_1$} to \texttt{Activity$_2$} in \texttt{Application 1}.
There are two implicit ICCs from \texttt{Activity$_2$} to \texttt{Activity$_3$} in \texttt{Application 1} and from \texttt{Activity$_2$} to \texttt{Activity$_4$} between \texttt{Application 1} and \texttt{Application 2}. 
Note that the target components of implicit ICC, \texttt{Activity$_3$} and \texttt{Activity$_4$}, have an \textit{Intent Filter} with the same action and category value as the \textit{Intent} used in \texttt{Activity$_2$}.
Each time there is an ICC, there may be a flow of data between components and potentially a privacy leak.

\subsection{FlowDroid}

FlowDroid~\cite{steven2014pldi} is a context-, flow-, field-, object-sensitive and lifecycle-aware static taint analysis tool for Android applications.
FlowDroid is based on Soot~\cite{lam2011soot} and Heros~\cite{Bodden2012}. 
The context-, flow-, field-, object-sensitives of FlowDroid are guaranteed by the precise call-graph of Soot and the IFDS/IDE~\cite{reps1995precise,srh96ide} based data-flow analysis of Heros.
A special main method, which considers all combinations of lifecycles, callbacks and entry points of Android components is generated to model data flows within the application. 
The sources and sinks used by FlowDroid are provided by SuSi~\cite{arzt2013susi}, also an open sourced tool used to fully automatically classify and categorize Android sources and sinks.
FlowDroid achieves 93\% recall and 86\% precision when detecting data leaks on DroidBench. 
FlowDroid has been mainly used on single component. 
However, with slight modifications, FlowDroid could also be used when multiple components are involved, i.e., for ICC analyses. 
Indeed, it's possible to use FlowDroid to compute paths for all individual components and then combines all those paths together, whatever there is a real link or not between these components. 
A major drawback is that this naive approach yields many false positives.

\subsection{Epicc}

Epicc~\cite{octeau2013effective} is a tool, also based on Soot and Heros, to identify ICC links.
In other words it finds links from ICC methods to their target components.
Epicc reduces the discovery of ICC in Android to an instance of the Interprocedural Distributive Environment (IDE) problem~\cite{srh96ide}. 
It uses data flow analysis to compute \textit{Intent} values at every ICC method call statements.
Experiments show that Epicc identifies 93\% of all ICC links and finds ICC vulnerabilities with far fewer false positives than the next best tool.

 \section{Motivating Example}
\label{example}

This section motivates our approach and illustrates the problem we solve through a concrete example. 
This example is detailed in Listing~\ref{motivatingExampleCode}, 
which presents code of \texttt{Application 1} introduced in Figure~\ref{fig:icc_graph}.
The app has three Activity components represented by \texttt{Activity$_1$}, \texttt{Activity$_2$} and \texttt{Activity$_3$} classes.
It also features \texttt{ButtonOnClickListener} a listener class used to handle button click events.
\texttt{Activity$_1$} registers a button listener for the \emph{to2} button (lines 6-11) and \texttt{Activity$_2$} registers one for the \emph{to3} button (line 15).

\begin{lstlisting}[caption={A Motivating Example Code},label=motivatingExampleCode]
//TelephonyManager telMnger; (default)
//SmsManager sms; (default)
class Activity1 extends Activity {
 void onCreate(Bundle state) {
  Button to2 = (Button) findViewById(to2a);
  to2.setOnClickListener(new OnClickListener(){
   String id = telMnger.getDeviceId();
   Intent i = new Intent(Activity1.this,Activity2.class);
   i.putExtra("sensitive", id);
   Activity1.this.startActivity(i);
  });}}
class Activity2 extends Activity {
 void onCreate(Bundle state) {
  Button to3 = (Button) findViewById(to3a);
  to3.setOnClickListener(new ButtonOnClickListener(this));
  Intent i = getIntent();
  String s = i.getStringExtra("sensitive");
  sms.sendTextMessage(number,null,s,null,null);
 }
 void onActivityResult(int,int,Intent){
  //log all the Extras of Intent
 }}
class Activity3 extends Activity {
 void onCreate(Bundle state) {
  Intent i = getIntent();
  String s = i.getStringExtra("sensitive");
  sms.sendTextMessage(number,null,s,null,null); 
 }}
class ButtonOnClickListener extends OnClickListener{
 //Activity act; (construct)
 void onClick(View view) {
  String id = telMnger.getDeviceId();
  Intent i = new Intent();
  i.setAction("test.ACTION"); //Action b
  i.putExtra("sensitive", id);
  act.startActivityForResult(i, 1);
 }}
\end{lstlisting}

When button \emph{to2} and \emph{to3} are clicked, the \texttt{onClick} method is executed and the user interface will change to \texttt{Activity$_2$} and to \texttt{Activity$_3$}, respectively.
In both cases, an \texttt{Intent} containing the device ID (lines 7 and 32), considered as sensitive data, is sent between two components by first attaching the data to the intent with the \texttt{putExtra} method (lines 9, and 35) and then by invoking either \texttt{startActivity} or \texttt{startActivityForResult} (lines 10 and 36).
Note that Listing \ref{motivatingExampleCode} exemplifies both the use of explicit and implicit intents.
At line 8, the intent is created by explicitly specifying the target class (\texttt{Activity}$_2$).
At line 34, only the intent action is specified with no explicit reference to the target.

In this example, \texttt{sendTextMessage} is directly executed when \texttt{Activity$_2$} or \texttt{Activity$_3$} is loaded since \texttt{onCreate} is the first method in the lifecycle of an \texttt{Activity}.
It sends the data retrieved from the \texttt{Intent} as a SMS to the specified phone number.

In this code, two privacy leaks occur: one when button \emph{to2} is clicked, the other when button \emph{to3} is clicked.
When  \emph{to2} is clicked, the device ID is transferred from \texttt{Activity$_1$} to \texttt{Activity$_2$} (line 10) and then \texttt{Activity$_2$} sends it outside the application (line 18). 

When  \emph{to3} is clicked, the device ID is transferred (line 36)  from \texttt{Activity$_2$} to \texttt{Activity$_3$}\footnote{As illustrated in Figure \ref{fig:icc_graph}, \texttt{Activity$_3$} has the appropriate \textit{Intent Filter} to catch the implicit \textit{Intent}}. Actually, the device ID (the source) is retrieved in class \texttt{ButtonOnClickListener} instantiated by \texttt{Activity$_2$}.
Finally, \texttt{Activity$_3$} sends the device ID outside the application (line 27).

The sensitive data leaks described above crosses two components: they cannot directly be detected since there is no real code connection between \texttt{startActivity} and \texttt{onCreate} (lines 10 and 13) or 
between \texttt{startActivityForResult} and \texttt{onCreate} (lines 36 and 24).  
Section~\ref{design} describes our approach to connect components to analyze paths between components and even between applications.

 \section{Definitions}
\label{definitions}

In order to better describe our approach, some android and taint analysis related concepts need to be defined.

\begin{figure}
\begin{center}
\begin{tikzpicture}
\tikzstyle{stmt} = [draw, circle, fill=white, inner sep = .04cm];
\path (0,0) node (first) {}
  ++(.2,0) node[stmt, label={[label distance=.3cm, above]\footnotesize Source}, label=below:$s_0$, fill=white] (source) {}
  ++(1,0) node[stmt, label=below:s$_1$] (s2) {} 
  ++(1,0) node[stmt, label=below:s$_2$] (s3) {} +(-.2, 0) node (first2) {}
  ++(1,0) node[stmt, label=below:s$_3$] (s4) {}
  ++(1,0) node[stmt, label=below:s$_4$] (s5) {} +(.2, 0) node (last2) {}
  ++(.5,0) node[] {$\dots$}
  ++(.5,0) node[stmt, label=below:s$_{11}$] (s11) {}
  ++(1,0) node[stmt, label=below:s$_{12}$, label={[label distance=.3cm, above]\footnotesize Sink}, fill=white] (sink) {}
  ++(.2,0) node (last) {}
;

\draw[-latex] (source) to[out=0, in=90] (s2);
\draw[-latex] (s2) to[out=0, in=90] (s3);
\draw[-latex] (s3) to[out=0, in=90] (s4);
\draw[-latex] (s4) to[out=0, in=90] (s5);
\draw[-latex] (s11) to[out=0, in=90] (sink);

\draw [decorate,decoration={brace,amplitude=10pt,mirror,raise=16pt},yshift=1cm]
(first) -- (last) node [black,midway, yshift=-35pt] {\footnotesize Tainted Path};

\draw [decorate,decoration={brace,amplitude=10pt,raise=15pt}]
(first2) -- (last2) node [black,midway, yshift=30pt] {\footnotesize Stmt Sequence};
\end{tikzpicture}
\end{center}
\caption{Representation of Statements, Source, Sink, Statement Sequence and Tainted Path.}
\label{fig:definitions}
\end{figure}
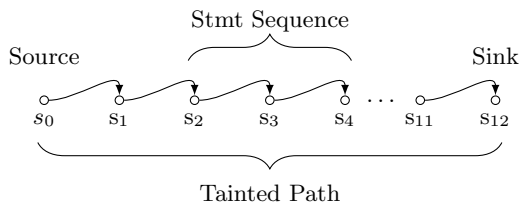

\emph{Control-Flow Graph (CFG)}
We detect data leaks by analyzing control-flow graphs of Android applications.
An application CFG consists of a collection of method CFGs linked together according to how they call one another.

\emph{Source Method.}
A source method returns data considered as private from the user's point of view into the application code.
For example, method \texttt{getDeviceId} (line 7\footnote{All the line numbers described in this section is referring to Listing~\ref{motivatingExampleCode}}) is a source method returning the device ID.

\emph{Sink Method.}
A sink method sends data out of the application.
For example, method \texttt{sendTextMessage} (line 27) is a sink method sending data to another phone using SMS.
We use sources and sinks computed for Android by the SuSi tool~\cite{arzt2013susi}.

\emph{ICC Method.}
An ICC method is used to trigger communication between two components.
For example, method \texttt{startActivity} (line 10) is an ICC method which triggers component communication from \texttt{Activity$_1$} to \texttt{Activity$_2$}.

\emph{Tainted Stmt.}
A tainted statement contains at least one tainted piece of data.
For example, \texttt{i.putExtra("sensitive ",id)} (line 9) is a statement containing the tainted data \texttt{id}.

\emph{Tainted Stmt Sequence.}
A tainted stmt sequence is a flow-sensitive sequence of tainted stmt.
For instance statements at line 9 and 10 form a tainted statement sequence.

\emph{Tainted Path.}
A tainted path is a tainted stmt sequence where
1) More than one stmt exist in the tainted path; 
2) The first stmt contains a source method; 
3) The last stmt contains a sink method.
Tainted Stmt, Tainted Stmt Sequence and Tainted Path are illustrated in Figure~\ref{fig:definitions}.

There are three types of tainted stmt paths in Android:
Intra-Component Communication, Inter-Component Communication (ICC) and Inter-Application Communication \linebreak (IAC) based tainted paths.

\begin{figure}
\begin{center}
 \begin{tikzpicture}\footnotesize
  \tikzstyle{layer} = [draw, rounded corners, minimum width=5cm, minimum height=.6cm];
  \tikzstyle{slayer} = [draw, rounded corners, minimum width=2cm];
  \path
  (0,0) node[layer] {ApkCombiner} +(0,0) node[layer, color=red, dashed, minimum width=5.2cm, minimum height=.8cm] {}
++(0,.8) node[layer] {Soot / Dexpler}
++(0,.8) node[layer] {Heros}
++(0,.8) node[layer, color=red, dashed, minimum width=5.2cm, minimum height=.8cm] {}
+(-1.5,0) node[layer, minimum width=1cm] {Epicc-IccTA} +(1.2,0) node[layer, minimum width=1cm] {FlowDroid-IccTA}
;
\end{tikzpicture}
\end{center}
\caption{The architecture of IccTA}
\label{fig:iccta_overview_a}
\end{figure}
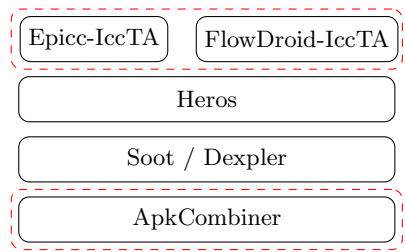

\begin{figure*}
\begin{center}
\begin{tikzpicture}\footnotesize
 \tikzstyle{apk} = [draw];
 \tikzstyle{data} = [draw, minimum height=1cm, minimum width=1.2cm, text width=1.2cm, align=center];
 \tikzstyle{alabel} = [midway, text width=2.4cm, align=center];

\path
 (0,0) node[data] (apk) {Apk}
++(3.5,0) node[data] (uj1) {\footnotesize Jimple}
++(7.5,0) node[data] (r1) {Tainted Paths}
;
 \path
(0,-1.5) node[data] (apkm) {Apk$^*$}
+(0,-1.2) node[data] (icc) {ICC Links}
+(-2.3, 0) node[inner sep=0, outer sep=0] (tmpa) {}
+(-3.5, .6) node[data] (apk1) {Apk$_1$}
+(-3.5, -.6) node[data] (apk2) {Apk$_2$}
+(-4.5, -1.4) node[inner sep=0, outer sep=0] (tmpc) {}
++(3.5,0) node[data] (uj2) {\footnotesize Jimple} 
+(0,-1.2) node[data] (iccdb) {Links\\DB}
+(1.2,-.6) node[inner sep=0, outer sep=0] (tmpb) {}
++(4,-.6) node[data, text width=1.7cm] (uj3) {\footnotesize Jimple \\+ ICC\phantom{.......} \\+ Lifecycle\\ + Callback}
++(3.5,0) node[data] (r2) {Tainted Paths}
;

\draw[-latex] (apk) -- (uj1) node[alabel] {\scriptsize (1.1) \\ Soot};
\draw[-latex] (uj1) -- (r1) node[alabel, yshift=.1cm] {\scriptsize (1.2)\\ FlowDroid \\ Analysis};

\draw[-] (apk1) -| (tmpa);
\draw[-] (apk2) -| (tmpa);
\draw[-latex] (tmpa) -- (apkm) node[alabel, yshift=.1cm] {\scriptsize (2.1) \\ Apk-\\Combiner};
\draw[-latex] (apkm) -- (uj2) node[alabel] {\scriptsize (2.2) \\ Soot};
\draw[-] (uj2) -| (tmpb);
\draw[-] (iccdb) -| (tmpb);
\draw[-latex] (tmpb) -- (uj3) node[alabel, yshift=.1cm] {\scriptsize (2.5) \\ FlowDroid-\\IccTA};
\draw[-latex] (uj3) -- (r2) node[alabel, yshift=.1cm] {\scriptsize (2.6)\\ FlowDroid \\ Analysis};
\draw[-latex] (apk2) |- (icc);
\draw[-latex] (icc) -- (iccdb) node[alabel] {\scriptsize (2.4)\\ Epicc-IccTA};
\draw[-] (apk1) -| (tmpc);
\draw[-latex] (tmpc) -- ([yshift=-2mm]icc.west) node[alabel, yshift=-.2cm] {\scriptsize (2.3) Epicc} 
node[alabel, pos=.65, yshift=.35cm] {\scriptsize (2.3) Epicc};
\end{tikzpicture}
\end{center}
\caption{Overview of IccTA (down) and FlowDroid (up).}
\label{fig:iccta_overview_b}
\end{figure*}

\emph{Intra-Component Tainted Path.}
An intra-component \linebreak tainted path is a tainted path only happening within a component.
In our motivating example, there is no intra-com\-ponent tainted path.
But if the \texttt{startActivity} call was replaced with a call to \texttt{sendTextMessage} which sends the  device id out of the application, there would be an intra-component tainted path (line 7-10).

\emph{ICC based Tainted Path.}
An ICC based tainted path is a tainted path among two or more components, i.e., there is at least one ICC method in the path.
In our motivating example, there is an ICC based tainted path from source method \texttt{getDeviceId} in \texttt{Activity$_1$} to sink method \texttt{sendTextMessage} in \texttt{Activity$_2$} through the \texttt{startActivity} ICC method (line 10).

\emph{IAC based Tainted Path.}
An IAC based tainted path is a tainted path between two or among more applications, i.e., it has at least one ICC method between two components of different applications.
There is no IAC based tainted path in our motivating example. 
But if the \texttt{Activity$_4$} in Figure~\ref{fig:icc_graph} sends the device id transferred from \texttt{Activity$_2$} out of the application, then there is an IAC based tainted path from \texttt{Application 1} to \texttt{Application 2}.

\emph{Privacy Leaks.}
If a tainted path is detected, it means that a privacy leak has been found.
In other words, some private data obtained from a \emph{source} method can flow through the tainted path to a \emph{sink} method.
 
 \section{IccTA}
\label{design}

In this Section we describe IccTA, our tool to detect privacy leaks in Android applications.
It uses static taint analysis to detect privacy leaks.
The main challenge for this is to solve the discontinuities problem introduced by the Android system.

We present the architecture of IccTA in Figure~\ref{fig:iccta_overview_a} where new or modified component are surrounded by a dashed line.
IccTA is the combination of Epicc-IccTA and FlowDroid-IccTA.
Epicc-IccTA relies on Epicc to incrementally compute ICC links from Android apps.
Both FlowDroid and Epicc are based on Soot~\cite{lam2011soot} and Heros~\cite{Bodden2012}. 
Soot is a framework to analyze Java based applications.
It uses the Dexpler~\cite{bartel:soap2012} plugin to convert Android Dalvik byte code to Soot's internal representation called Jimple and relies on Spark~\cite{spark} to build accurate call graphs.
Heros is a scalable implementation of IFDS~\cite{rhs95ifds} and IDE~\cite{srh96ide}, two frameworks to perform data flow analysis.
Analyzing multiple applications is done using ApkCombiner.
It combines multiple apps to a single one to ease the analysis of IccTA.

Figure~\ref{fig:iccta_overview_b} is a comparison between IccTA and FlowDroid.
FlowDroid (\ref{fig:iccta_overview_b} up) first converts the Android bytecode to Jimple in step (1.1).
Then, in step (1.2), it analyzes the Jimple code to detect tainted paths in single Android components.

IccTA (\ref{fig:iccta_overview_b} down) can analyze one or multiple Android applications.
If more than one application is analyzed, it uses ApkCombiner to merge the Android applications in a single application in step (2.1).
The Android application's bytecode is then converted to Jimple in step (2.2).
In parallel, Epicc-IccTA analyzes all the input applications (Apk$_1$ and Apk$_2$ in the Figure) to generate ICC Links in step (2.3) and stores the results to a database in step (2.4).
IccTA uses ICC links generated by Epicc-IccTA to connect Android components in the Jimple code in step (2.5).
Steps (2.2) and (2.6) correspond to FlowDroid's steps (1.1) and (1.2): the Jimple code is updated to take into account lifecycles and callbacks of components and the taint analysis is launched to generate a list of tainted paths.

\subsection{FlowDroid-IccTA: Reducing the ICC pro-blem to an Intra-Component problem}
\label{subsec:flowdroid_iccta}

Since there is no direct code connection between two Android components, FlowDroid cannot detect ICC based privacy leaks with precision.
In this section, we describe how FlowDroid-IccTA reduces the ICC problem to an intra-\linebreak component problem on which FlowDroid can perform an highly precise data-flow analysis.
Our approach instruments the \emph{Jimple} code of Android applications to connect components directly in the code.

As mentioned in the introduction, there are three types of discontinuities in Android: (1) ICC methods, (2) lifecycle methods and (3) callback methods.
We first describe how FlowDroid-IccTA tackles ICC methods in Section~\ref{subsucsec:icc_methods}.
Then, we detail how FlowDroid-IccTA resolves lifecycle and callback methods in Section~\ref{subsubsec:lifecycle_callback_methods}.
Finally, using our motivating example of Listing~\ref{motivatingExampleCode}, we illustrate the code instrumentation process in Section~\ref{subsubsec:an_example}.

\subsubsection{ICC Methods}
\label{subsucsec:icc_methods}

\begin{figure}
\centering
\begin{tikzpicture}
\tiny

\path (0, 0) node[label=left:(A)] {
 \begin{lstlisting}[basicstyle=\tiny\ttfamily, numbers=none,]
  // modifications of Activity1
  Activity1.this.startActivity(i);
  IpcSC.redirect0(i);
\end{lstlisting}
}
+(-.22\columnwidth, .11) node (l) {}
+( .22\columnwidth, .11) node (r) {}
++(0,-1.5) node [label=left:(B)] {
 \begin{lstlisting}[basicstyle=\tiny\ttfamily, numbers=none,]
// creation of a helper class
class IpcSC {
 static void redirect0(Intent i) {
  Activity2 a2 = new Activity2(i);
  a2.dummyMain();
 }
}
\end{lstlisting}
}
++(4.2,+.6) node [label=left:\scriptsize(C)] {
 \begin{lstlisting}[basicstyle=\tiny\ttfamily, numbers=none,]
// modifications in Activity2
public Activity2(Intent i) {
  this.intent_for_ipc = i;
}
public Intent getIntent() {
 return this.intent_for_ipc;
}
public void dummyMain() {
 // lifecycle and callbacks
 // are called here
}
\end{lstlisting}
}
;
\path[draw] (l) -- (r);

\end{tikzpicture}
\caption{Code Modifications to Handle ICC Communication between Activity$_1$ and Activity$_2$.
The \texttt{startActivity} ICC method is replaced (A) by a call to code that instantiates and calls the ``main'' method of Activity$_2$ (B).
The target component class is updated to handle Intent objects directly, by modeling the Android system behavior (C).
}
\label{fig:icc_after_instrumentation}
\end{figure}

As shown in Figure~\ref{fig:iccta_overview_b}, the ICC problem is solved at step 2.5.
This is where the \emph{Jimple} code is updated by FlowDroid-IccTA to connect components.
This code modification is required for all ICC methods (listed in Table \ref{tab:icc_methods}). 
We detail these modifications for the two most used ICC methods: \texttt{startActivity} and \texttt{startActivityForResult}.
We handle ICC methods for \textit{Service}s and \textit{Broadcast Receiver}s in a similar way.

\textbf{StartActivity.}
Figure~\ref{fig:icc_after_instrumentation} shows the code transformation done by FlowDroid-IccTA for the ICC link between \texttt{Activity$_1$} and \texttt{Activity$_2$} of our motivating example.\linebreak
FlowDroid-IccTA first creates a helper class named \texttt{IpcSC} (B in Figure~\ref{fig:icc_after_instrumentation}) which acts as a bridge connecting the source and destination components.
Then, the \texttt{startActivity} ICC method is removed and replaced by a statement calling the generated helper method (\texttt{redirect0}) (A).

In (C), FlowDroid-IccTA generates a constructor method taking an \texttt{Intent} as parameter, a \texttt{dummyMain} method to call all related methods of the component (i.e., lifecycle and callback methods) and overrides the \texttt{getIntent} method.
An Intent is transferred by the Android system from the caller component to the callee component.
We model the behavior of the Android system by explicitly transferring the Intent to the destination component using a customized constructor method, \texttt{Activity$_2$(Intent i)}, which takes an \texttt{Intent} as its parameter and stores the Intent to a newly generated field \texttt{intent\_for\_ipc}.
The original \texttt{getIntent} method asks the Android system for the incoming Intent object.
The new \texttt{getIntent} method models the Android system behavior by returning the Intent object given as parameter to the new constructor method.

The helper method \texttt{redirect0} constructs an object of type \texttt{Activity$_2$} (the target component) and initializes the new object with the \texttt{Intent} given as parameter to the helper method.
Then, it calls the \texttt{dummyMain} method of \texttt{Activity$_2$}.

To resolve the target component, i.e., to automatically infer what is the type that has to be used in the method \texttt{redirect0} (in our example, to infer \texttt{Activity$_2$}),  
Flowdroid-IccTA uses the ICC links computed by Epicc-IccTA.
Epicc-IccTA resolve the target component not only for explicit  \emph{intents}, but also for implicit  \emph{intents}. Therefore, there is no difference for Flowdroid-IccTA to handle explicit or implicit \emph{intent} based ICCs.

\textbf{StartActivityForResult.}
There are some special ICC methods in Android, such as \texttt{startActivityForResult}.
A component $C_1$ can use this method to start a component $C_2$.
Once $C_2$ finishes running, $C_1$ runs again with some result data returned from $C_2$.
The control-flow mechanism of \texttt{startActivityForResult} is shown in Figure~\ref{fig:mechanism_startActivityForResult}.
There are two discontinuities: one from (1) to (2), similar to the discontinuity of the \texttt{startActivity} method, and the other from (3) to (4).

The \texttt{startActivityForResult} ICC method has a more complex semantic compared to common ICC methods that only trigger one-way communication between components (e.g., \texttt{startActivity}).
Figure~\ref{fig:instrumenting_startActivityForResult} shows how the code is instrumented to handle the \texttt{startActivityForResult} method in our motivating example.
To stay consistent with common ICC methods, 
we do not instrument the \texttt{finish} method of \texttt{Activity$_3$} to call \texttt{onActivityResult} method.
Instead, we generate a field \texttt{intent\_for\_ar} to store the \textit{Intent} which will be transferred back to \texttt{Activity$_2$}.
The \textit{Intent} that will be transfered back is set by the \texttt{setResult} method.
We override the \texttt{setResult} method to store the value of \textit{Intent} to \texttt{intent\_for\_ar}.
The helper method \texttt{IpcSC.redirect0} does two modifications to link these two components directly.
First, it calls the \texttt{dummyMain} method of destination component.
Then, it calls the \texttt{onActivityResult} method of the source component.

\vspace{-3mm}
\begin{figure}[H]
\begin{center}
\begin{tikzpicture}
 \tikzstyle{nodeS} = [draw, minimum width=2.6cm, minimum height=.6cm, font=\scriptsize]
 \tikzstyle{nbr} = [solid, midway, above, circle, draw, inner sep=.12mm, yshift=2mm, font=\scriptsize];
 \tikzstyle{emptyS} = [inner sep=0cm, outer sep=0cm]

\path
(0, 0)node[nodeS] (a1a) {Activity$_2$ Entry Point}
++(0, -1cm) node[nodeS] (a1b) {startActivityForResult}
+(0, -.5cm) node[draw, dotted, minimum width=3.1cm, minimum height=4cm, rounded corners, label=above:\footnotesize Activity$_2$] {}
++(0, -2cm) node[nodeS] (a1c) {onActivityresult}

++(2.6cm, 1cm) node [draw, circle, outer sep = 0cm, inner sep = .0cm, double, text width=1.2cm, text centered] (as) {Android\\ System}

++(2.6cm, 2cm) node[nodeS] (a2a) {Activity$_3$ Entry Point}
++(0, -1cm) node[nodeS] (a2b) {setResult}
+(0, -.5cm) node[draw, dotted, minimum width=3.1cm, minimum height=4cm, rounded corners, label=above:\footnotesize Activity$_3$] {}
++(0, -2cm) node[nodeS] (a2c) {finish}
;

\draw[-latex] (a1a) -- (a1b);
\draw[-latex] (a2a) -- (a2b);
\draw[-latex] (a2b) -- (a2c);

\draw[-latex, dashed] (a1b) to[out=-90, in =170] node[nbr] {1} (as) ;
\draw[-latex, dashed] (as) to[out=30, in =180] node[nbr, left, xshift=-.1cm] {2} (a2a);
\draw[-latex, dashed] (a2c) to[out=180, in =-30] node[nbr, xshift=.2cm] {3} (as);
\draw[-latex, dashed] (as) to[out=-170, in =90] node[nbr, below, yshift=-.4cm] {4} (a1c);
\end{tikzpicture}
\end{center}
\caption{The control-flow mechanism of \texttt{startActivityForResult}}
\label{fig:mechanism_startActivityForResult}
\end{figure}

\vspace{-5mm}

\subsubsection{Lifecycle and Callback Methods}
\label{subsubsec:lifecycle_callback_methods}

\begin{figure}
\centering
\begin{tikzpicture}
\tiny

\path (0, 0) node[label=left:(A)] {
 \begin{lstlisting}[basicstyle=\tiny\ttfamily, numbers=none,]
act.startActivityForResult(i); 
IpcSC.redirect0(act, i);
\end{lstlisting}
}
+(-.2\columnwidth, .21) node (l) {}
+( .2\columnwidth, .21) node (r) {}
++(0,-1.2) node [label=left:(C)] {
 \begin{lstlisting}[basicstyle=\tiny\ttfamily, numbers=none,]
void setResult(Intent i) {
 this.intent_for_ar = i;
 a2.dummyMain();
}
public Intent getIntentFAR() {
 return this.intent_for_ar;
}
\end{lstlisting}
}
++(4.2,+.5) node [label=left:(B)] {
 \begin{lstlisting}[basicstyle=\tiny\ttfamily, numbers=none,]
class IpcSC {
 static void redirect0(Activity a2, 
                Intent i) {
  Activity3 a3 = new Activity3(i);
  a3.dummyMain();
  Intent retI = a3.getIntentFAR();
  a2.onActivityResult(retI);
 }
} 
\end{lstlisting}
}
;
\path[draw] (l) -- (r);

\end{tikzpicture}
\caption{An Example about running FlowDroid-IccTA to \texttt{startActivityForResult} ICC method. (A) represents the modified code of \texttt{ButtonOnClickListener} and (C) the modified code of \texttt{Activity$_3$}. (B) is the glue code connecting \texttt{ButtonOnClickListener} and \texttt{Activity$_3$.}
Some method parameters are not represented to simplify the code.
}
\label{fig:instrumenting_startActivityForResult}
\end{figure}

One challenge when analyzing Android applications is to tackle the callback methods and the lifecycle methods of components.
There is no direct call among those methods in the code of applications since the Android system handles lifecycles and callbacks.
For callback methods, we need to take care of not only the methods triggered by the User Interface (UI) events (e.g., \texttt{onClick}) but also of callbacks triggered by Java or the Android system (e.g., the \texttt{onCreate} method).
In Android, every component has its own lifecycle methods.
To solve this problem, IccTA generates a \texttt{dummyMain} method for each component in which we model all the methods mentioned above so that our CFG based approach is aware of them.
Note that FlowDroid also generates a \texttt{dummyMain} method, but it is generated for the whole app instead of for each component like we do.

\subsubsection{The CFG of instrumented motivating example}
\label{subsubsec:an_example}

\begin{figure*}[!t]
\centering
 \begin{tikzpicture}\scriptsize
\tikzstyle{lbl} = [midway, align=center, color=black];
\tikzstyle{doubleS} = [-latex, double, color=black!30!green];
\tikzstyle{dashedS} = [-latex, dashed, color=blue];
\tikzstyle{dottedS} = [-latex, dotted, color=black];
\tikzstyle{normalS} = [-latex, color=black];

\newdimen\xsep
\xsep=1cm
  \path (0,0) node (1) {String id = telMnger.getDeviceId();}
++(0, -\xsep) node (2) {i.putExtra("sensitive", id);}
++(0, -\xsep) node (3) {ipcSC.redirect0(i);}
++(0, -\xsep) node (4) {\color{blue}return-site;}

++(4, 2.5\xsep) node (5) {Activity2 a2 = new Activity2(i);}
++(0, -\xsep) node (6) {\color{blue}return-site;}
++(0, -\xsep) node (7){a2.dummyMain();}
++(0, -\xsep) node (8) {\color{blue}return-site;}

++(4, 3.5\xsep) node (9) {this.intent\_for\_ipc = i;}
++(0, -\xsep) node {}
++(0, -\xsep) node (10) {onCreate(null);}
++(0, -\xsep) node (11) {\color{blue}return-site;}
++(0, -\xsep) node {}

++(4, 4\xsep) node {}
+(1, 0) node (12) {return this.intent\_for\_ipc;}
++(0, -.6\xsep) node (13) {Intent i = getIntent();}
++(0, -\xsep) node (14) {\color{blue}return-site;}
++(0, -\xsep) node (15) {String s = i.getStringExtra("sensitive");}
++(0, -\xsep) node (16) {sendTextMessage(s);}
;

\path (5.7, -4.2\xsep) node[minimum width=.88\textwidth, minimum height=.6cm, draw, dotted, rounded corners] {}
+(0,-.5) node {};
\path (0, -4.2\xsep) node[] {}
+(-2, 0) node (a1) {}
+(-1, 0) node (a2) {}
+(0, 0) node {normal edge};
\path (4, -4.2\xsep) node[] {}
+(-2, 0) node (b1) {}
+(-1, 0) node (b2) {}
+(.2, 0) node {call-to-start edge};
\path (8, -4.2\xsep) node[] {}
+(-2, 0) node (c1) {}
+(-1, 0) node (c2) {}
+(.2, 0) node {call-to-return edge};
\path (12, -4.2\xsep) node[] {}
+(-2, 0) node (d1) {}
+(-1, 0) node (d2) {}
+(.2, 0) node {exit-to-return edge};
\draw[normalS] (a1) -- (a2);
\draw[doubleS] (b1) -- (b2);
\draw[dottedS] (c1) -- (c2);
\draw[dashedS] (d1) -- (d2);

\draw[normalS] (1) -- (2) node[lbl, xshift=.3cm] {(1)};
\draw[normalS] (2) -- (3) node[lbl, xshift=.3cm] {(2)};
\draw[dottedS] (3) -- (4);
\draw[dottedS] (5) -- (6);
\draw[normalS] (6) -- (7) node[lbl, xshift=.3cm] {(6)};
\draw[dottedS] (7) -- (8);
\draw[dottedS] (10) -- (11);
\draw[dottedS] (13) -- (14);
\draw[normalS] (14) -- (15) node[lbl, xshift=.3cm] {(11)};
\draw[normalS] (15) -- (16) node[lbl, xshift=.3cm] {(12)};

\draw[doubleS] (3) to[out=0, in=-170, distance=1.5cm] node[lbl, xshift=.3cm] {(3)} (5) ;
\draw[doubleS] (5) to[out=0, in=-180, distance=.9cm] node[lbl, xshift=-.4cm, yshift=.3cm] {(4)} (9);
\draw[doubleS] (7) to[out=0, in=-180, distance=1.5cm] node[lbl, xshift=.3cm] {(7)} (10);
\draw[doubleS] (10) to[out=0, in=-180, distance=1.5cm] node[lbl, xshift=.3cm] {(8)} (13);
\draw[doubleS] (13) to[out=150, in=-180, distance=.6cm] node[lbl, xshift=-.3cm] {(9)} (12);

\draw[dashedS] (12) to[out=-20, in=-0, distance=1.2cm] node[lbl, xshift=.4cm] {(10)} (14);
\draw[dashedS] (8) to[out=180, in=-0, distance=1.5cm] (4);
\draw[dashedS] (9) to[out=-90, in=-0, distance=1.1cm] node[lbl, xshift=-.3cm, yshift=.2cm] {(5)} (6);
\draw[dashedS] (11) to[out=180, in=-0, distance=1.5cm] (8);
\draw[dashedS] (16) to[out=180, in=-0, distance=1.5cm] (11);
 \end{tikzpicture}
\caption{The control-flow graph of the instrumented motivating example}
\label{fig:icc2ipc}
\end{figure*}
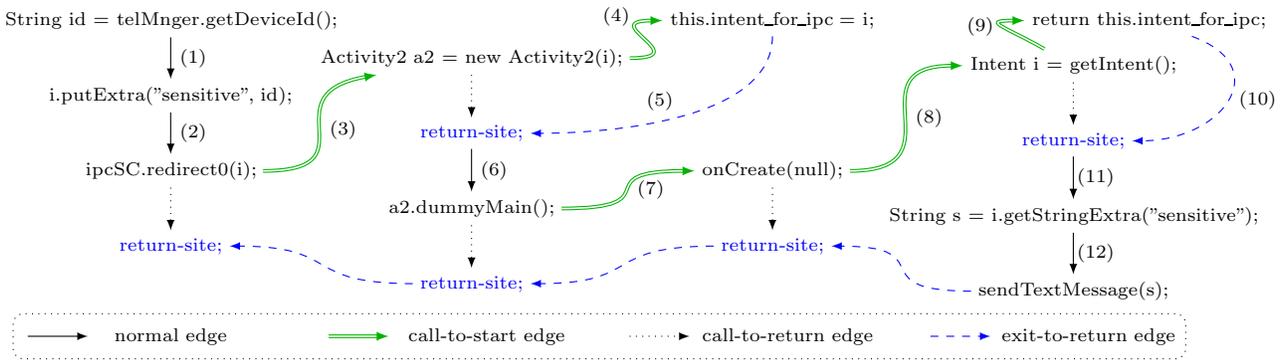

Figure~\ref{fig:icc2ipc} represents the CFG of the instrumented motivating example presented in Listing~\ref{motivatingExampleCode}.
In the CFG, \texttt{getDeviceId} is a \emph{source} method in the anonymous \texttt{OnClickListener} class (line 6) called by \texttt{Activity$_1$}.
Method \texttt{sendTextMessage} is a \emph{sink} in \texttt{Activity$_2$}.
There is an intra-component tainted statement path from the \emph{source} method to \emph{sink} method (represented by edges 1 to 12).

Figure~\ref{fig:icc2ipc} also shows that IccTA builds a precise cross-component control-flow graph. 
Since we use an technique instrumenting the code to build the CFG, the context of a static analysis is kept between components.
This enables IccTA to analyze data-flows between components and thereby enables IccTA to have a better precision than existing approaches.  

\subsection{ApkCombiner: Reducing an IAC problem to an ICC problem}
\label{subsec:apk_combiner}

In Android, Inter-Application Communication (IAC) is similar to Inter-Component Communication (ICC).
Indeed, IAC also relies on component communication, 
except that the source component and the destination component belong to different applications.
If we can connect applications, an \textit{IAC Problem} becomes a standard \textit{ICC Problem}. 

\textbf{Analyzing Multiple Applications.}
As shown in Figure~\ref{fig:iccta_overview_b}, FlowDroid can only analyse one application at a time.
Therefore, we develop a tool, \emph{ApkCombiner}, to combine multiple apps into one.
ApkCombiner combines all the parts of Android apps including bytecodes, assets, manifest and all the resources.
Then, we use IccTA to analyze the combined app to compute IAC based privacy leaks.
As FlowDroid-IccTA handles the combined application as a single applications, it only detects ICC based privacy leaks.
To distinguish ICC leaks from IAC leaks, IccTA checks if 
all statements of the tainted path belong to the same application or not.

\textbf{Reducing the Number of Combined Apps to Analyze.}
In practice, when increasing the number of applications to analyze, and if all those applications are combined with ApkCombiner, the processing time and memory requirement of FlowDroid-IccTA also grows. 
To solve this problem, we need to decrease the number of Android apps to combine.
Our solution is to  build an IAC graph, where a node is an application and an edge a link, to represent the dependencies between applications.
The idea behind being that if there is no link between two applications there is no need to combine them.

The IAC graph is made up of small independent IAC (sIAC) graphs (connected components).
Given a sIAC graph, ApkCombiner combines all the nodes (apps) in it into one app, then IccTA extracts leaks from the resulting app.
However, in some case, if a sIAC graph still contains a lot of nodes.
This will also limit our approach to be scalable.
Our solution is to limit the length (how many apps are involved) of an IAC leak\footnote{In practice we have not seen a leak going through more than 2 apps.}.
For example, If a sIAC graph contains 10 nodes (where \texttt{A$_i$} is connected to \texttt{A$_{i+1}$, $i \in \{1, 9\}$}) and the length limitation is set to five.
Then, the sIAC graph is split into five sIACs (e.g., one sIAC is from \texttt{A2} to \texttt{A6}) that IccTA can analyze.
The trade-off limitation length enables our approach to become scalable.

Another good point of building an IAC graph is that new applications can be added to the graph in an iterative and incremental manner.
When new apps are involved, we only run them against Epicc-IccTA and add them to the existing IAC graph. 
We do not need to run the previously computed apps again when adding the new apps to the IAC graph.

In short, by building an IAC graph, the original set of Android applications is split into multiple small sets that IccTA can analyze.

 \section{Evaluation}
\label{evaluation2}

Our evaluation addresses the following research questions:
\begin{description}
\item[RQ1] How does IccTA compare to commercial taint-analysis tools for Android and FlowDroid in terms of precision and recall?
\item[RQ2] Can IccTA find leaks in real-world applications and how fast is it?
\item[RQ3] How do IccTA compare to other academic ICC leak detection approaches?
\end{description}

\subsection{RQ1: IccTA vs FlowDroid and Commercial Tool}
\label{subsec:droidbench}
 
We evaluate and compare IccTA with FlowDroid and IBM AppScan Source 9.0 on DroidBench to test for ICC and IAC leaks.
Unfortunately, we were unable to compare IccTA to other static analysis tools as their authors did not make them available.

\textbf{DroidBench.}
DroidBench~\cite{droidbench}  is a set of hand crafted Android applications for which all leaks are known in advance.
The fact of knowing all leaks in the applications is called the \emph{ground truth} and is used to evaluate how well static and dynamic security tools find data leaks. 
DroidBench version 1.2 contains 64 different test cases with different privacy leaks. 
However, all the leaks in DroidBench are intra-component privacy leaks.
Thus, we developed \numberOfDroidbenchApps{} apps and 23 test cases to extend DroidBench with ICC and IAC leaks.
A test case is applied on one application to test for ICC and on two applications to test for IAC.
In total, 18 apps contain inter-component privacy leaks and 6 apps contain inter-app privacy leaks.
The new set of test cases  covers each of the top \topICCMethods{} ICC methods in Table \ref{tab:icc_methods}.
Moreover, among the 26 new apps, two of them do not contain any privacy leaks.
If a tool detects privacy leaks on these two apps, the detected leaks are false alarms.
Finally, for each test case application we add an unreachable component containing a sink.
These unreachable components are used to flag tools that do not properly construct links between components.
 
The 23 test cases are listed in the first column of Table~\ref{tab:droidbench}.

\textbf{IccTA.}
We run IccTA on all the 23 test cases. 
The results are shown in Table~\ref{tab:droidbench}.
IccTA successfully passes 18 test cases, with 17 test cases containing 19 privacy leaks and one test case (\texttt{startActivity5}) with no leak.

Among the detected privacy leaks, three of them are IAC based privacy leaks and the remaining ones are ICC based privacy leaks.
In the \texttt{startActivity5} test case, the source component uses an implicit intent with data type \emph{text/plain} to start another activity.
However, no other activity in this test case declares that it can receive an intent with data type \emph{text/plain}.
That means there is no connection among the components in \texttt{startActivity5} test case.
As IccTA takes into consideration the data type of an intent it does not report any privacy leak for this test case.

The \texttt{startActivity4} test case does not contain any leaks. 
However, IccTA does report a false warning.
The reason is that the source component uses an implicit intent with an URI to start another activity.
Since IccTA relies on Epicc which does over-approximate URIs links, it reports a false leak.

The current version does not take into account \texttt{Content Provider}s.
This is why IccTA misses leaks for the \texttt{insert1}, \texttt{delete1}, \texttt{update1}, and \texttt{query1} test cases.
All the four test cases are related to \texttt{Content Provider}.

\textbf{FlowDroid.}
FlowDroid has been evaluated on the first version of DroidBench in \cite{steven2014pldi}. 
In table~\ref{tab:droidbench}, we present the results of FlowDroid on the new 23 test cases. As already explained, FlowDroid has been initially proposed to detect leak in single Android component. 
However, we can use FlowDroid in a way that it computes paths for all individual components and then combines all those paths together (whatever there is a real link or not). As a result, we expect that FlowDroid detects most of the leaks but yields several false positives.   
Results of Table \ref{tab:droidbench} confirm this expectation: FlowDroid has a high recall (69.6\%) and a low precision (23.9\%).
FlowDroid misses three more leaks than IccTA in \texttt{bindService\{2,3,4\}}. 
After investigation, we discover that FlowDroid does not consider some callback methods for service components.

\textbf{AppScan.}
AppScan Source 9.0 requires a lot of manual initialization work since it has no default sources/sinks configuration file and is unable to analyze Android applications without specifying the entry points of every components.
We define the \texttt{getDeviceId} and \texttt{log} methods, that we always use in DroidBench for ICC and IAC leaks, as source and sink, respectively.
We also add all components' entry point methods (such as \texttt{onCreate} for activities) as callback methods so AppScan knows where to start the analysis.
AppScan is natively unable to detect inter-component data-flows and only detects intra-component flows.
AppScan has the same drawbacks as FlowDroid and should have a high recall and low precision on DroidBench.
We use an additional script to combine the flows between components.
As expected AppScan's recall is high (56.5\%) and its precision low (21.0\%).
Compared to FlowDroid, AppScan does worse.
Indeed, AppScan does not correctly handle \texttt{startActivityForResult} and thus misses leaks going through methods receiving results from the called activities in \texttt{startForResult\{2,3,4\}}.

\textbf{Conclusion.}
IccTA outperforms both the commercial taint-analysis tool AppScan 9.0 and FlowDroid in terms of precision and recall.

\begin{table}[!t]
\centering
\footnotesize 
\caption{DroidBench test results}

\hit = correct warning, \fp = false warning, \fn = missed leak \\
multiple circles in one row: multiple leaks expected\\
all-empty row: no leaks expected, none reported\\
$^\dagger$ C/A: \# of Components / \# of Applications\\

\label{tab:droidbench}
\begin{tabular}{ 
 >{\columncolor[gray]{.9}[6pt]}l
p{1.55cm} 
 >{\columncolor[gray]{.9}[6pt]} p{1.25cm} 
p{.9cm}
}
\hline
\rowcolor{black!30}
\textbf{Test Case} (\textbf{C/A})$^\dagger$ & \textbf{FlowDroid} & \textbf{AppScan} & \textbf{IccTA} \\ 
\hline
\hline
\multicolumn{4}{c}{\cellcolor{black!20}Inter-Component Communication} \\ 
startActivity1 (3/1) 	& 	\hit \fp				&\hit \fp			& \hit  \\ 
startActivity2 (4/1) 	&  	\hit (4\fp) 			&\hit (4\fp)		& \hit  \\  
startActivity3 (6/1) 	&  	\hit (32\fp)			&\hit (32\fp)		& \hit  \\  
startActivity4 (3/1) 	&  	\fp \fp				&\fp \fp		& \fp   \\  
startActivity5 (3/1) 	&  	\fp \fp				&\fp \fp		& 	\\  
startForResult1 (3/1) 	&  	\hit					&\hit		& \hit  \\  
startForResult2 (3/1) 	&  	\hit					&\fn			& \hit  \\  
startForResult3 (3/1) 	&  	\hit \fp				&\fn			& \hit  \\ 
startForResult4 (3/1) 	&  	\hit	 \hit \fp		&\hit \fn	& \hit \hit\\  
startService1 (3/1) 		&  	\hit \fp				&\hit \fp	& \hit \\ 
startService2 (3/1) 		&  	\hit \fp				&\hit \fp		& \hit \\ 
bindService1 (3/1) 		&  	\hit \fp				&\hit \fp	& \hit	   \\
bindService2 (3/1) 		&  	\fn					&\fn		& \hit	\\ 
bindService3 (3/1) 		&  	\fn					&\fn		& \hit  \\ 
bindService4 (3/1) 		&  	\hit \fp \fn			&\hit \fp \fn	& \hit \hit \\ 
sendBroadcast1 (3/1) 	& 	\hit	 \fp				&\hit \fp		& \hit  \\
insert1	(3/1) 			&	\fn					&\fn		& \fn   \\
delete1	(3/1) 			&	\fn					&\fn		& \fn   \\
update1	(3/1) 			&	\fn					&\fn		& \fn   \\
query1 (3/1) 			&	\fn					&\fn		& \fn   \\
\multicolumn{4}{c}{\cellcolor{black!20}Inter-App Communication} \\ 
startActivity1 (4/2) 	&	\hit \fp				& \hit \fp		& \hit \\  
startService1 (4/2) 		&  	\hit \fp				& \hit \fp		& \hit \\  
sendBroadcast1 (4/2) 	&  	\hit \fp				& \hit \fp		& \hit \\

\multicolumn{4}{c}{\cellcolor{black!20}Sum, Precision, Recall and F$_1$} \\ 
\hit, higher is better 				& 16 		& 13			& 19 \\ 
\fp, lower is better					& 51			& 49			& 1 \\ 
\fn, lower is better 				& 7			& 10			& 4 \\
Precision $\hit / (\hit + \fp)$ 		& 23.9\% 	& 21.0\%		& \precisionDroidbench{} \\  
Recall $\hit / (\hit + \fn)$    		& 69.6\% 	& 56.5\%		& \recallDroidbench{} \\ 
F$_1$ $2\hit / (2\hit + \fp + \fn)$	& 0.36 		& 0.31		& 0.88\\  
\hline

\end{tabular}
\end{table}

\subsection{RQ2: IccTA and Real-World Apps}
\label{subsec:experimental_results}

We run the experiments on a Core i7 CPU running a Java VM with 8 Gb of heap.
To evaluate our approach, we use IccTA to analyze 3000 Android apps downloaded from the Google Play market as well as some third-party markets (e.g., wandoujia). 
IccTA process 3000 apps in about 100 hours.
IccTA does not detect any leak for 2575 (85.83\%) applications.
IccTA reports 425 applications containing privacy leaks.
Among the 425 apps, 411 apps only contain intra-component leaks and 14 apps contain at least one ICC leak.
From those 14 apps, 13 contain both intra-component leaks and ICC leaks.
IccTA detects 6989 IAC links.
Among those IccTA detects one IAC leak.
This result indicates that components do communicate and share data, but it is rare that an inter-application leak occurs.

For intra-app leaks, IccTA detects 5986 leaks in the 425 apps.
Among the detected leaks, 147 (2.5\%) are ICC privacy leaks.
We manually check the 147 reported ICC leaks and found out that 17 (11.6\%) are false positives.
In other words, IccTA achieves a precision of 88.4\% on real-word apps. 
The false positives comes from Epicc that generates false positives for links between components.

We summarize the frequently used \emph{source} methods and \emph{sink} types (Java classes) in Table~\ref{tab:experimental_results_top_5} from the 425 apps having at least one leak.
Note that we only count such \emph{source} and \emph{sink} methods that appear in the detected leaks.
The most used \emph{source} method is \texttt{openConnection} and it is used 601 times in 169 apps.
The most used \emph{sink} types is \texttt{Log} and it is used 2755 times in 261 apps.
The reason why we study \emph{sink} types instead of \emph{sink} methods is that there are a lot of \emph{sink} methods in a same \emph{sink} type.
Take the Log \emph{sink} type as an example, there are eight \emph{sink} methods which log the private data to disk.

\begin{table}[!t]
\centering
\footnotesize
\definecolor{LightGray}{gray}{0.9}
\caption{The top 5 used source methods and sink types}
\label{tab:experimental_results_top_5}
\begin{tabular}{p{2.9cm} c p{2.5cm}}
\hline
\rowcolor{black!30}
\textbf{Method/Type}		&	\textbf{Counts(\#.)}  	& \textbf{Detail} \\ 
\hline
\hline

\rowcolor{black!20}
\multicolumn{3}{c}{Source Methods}  \\  
openConnection 			&	601 			& http connection\\ 
getLongitude 			&	514 			& longitude\\ 
getLastKnownLocation 	&	448 			& Location \\
getDeviceId 				&	403 			& IMEI or ESN \\ 
getCountry 				&	265 			& country code\\ 

\rowcolor{black!20}
\multicolumn{3}{c}{Sink Types}  \\ 
Log 					&	2755  	& error or warn	\\ 
URL 					&	821 		& execute  \\ 
SharedPreferences 	&	717  	& putInt, putString \\
Message 				&	339 		& sendTextMessage\\ 
File 				&	9  	& write(string)\\ 
\hline
\end{tabular}
\end{table}

Let us describe in details three leaks, one for each type of leak.

\textbf{Intra-component leak: }\texttt{\emph{bz.prana.myphonelocator.}}
I\-ccTA detects an intra-component privacy leak starting from the \texttt{getLongitude} source method in method \texttt{onLocationChanged} of class \texttt{.SMSReceiver\$MyLocationListener\footnote{The package name is omitted when the class name starts with the package name}}.
The location is sent out of the app through SMS by the \texttt{sendTextMessage} \emph{sink} method in method \texttt{smsReply} of class \texttt{.SMSReceiver}.
The app is designed to send the location outside the device through SMS.
However, to distinguish the intention of detected privacy leaks is out of scope of this paper.
We take it as our further work.

\textbf{ICC leak: }\texttt{\emph{com.dikkar.ifind.}}
An ICC based privacy leak is detected by IccTA on this application.
In method \texttt{onLocationChanged} of class \texttt{.iFindPlaces}, the \texttt{getLongitude} \emph{source} method is called and returns the location of the Android phone. 
Then, the location is transferred to another component named \texttt{.PlaceDetail}, where method \texttt{b} of class \texttt{j} is called.
In method \texttt{b}, a \emph{sink} method \texttt{Log.d} logs the location into disk with \texttt{ServiceHandler} tag name.
To verify the detected leaks, we developed an Android application named LogParser.
By giving the permission \texttt{android.permission.READ\_LOGS}
\footnote{Starting from Android 4.1 it is no more granted to regular apps, but it can still be granted to either vendor apps or apps running on rooted phones.},
LogParser reports all the locations logged by \texttt{Find Places}.

\textbf{IAC leak: }\texttt{\emph{com.bi.mutabaah.id to jp.benishouga.cl\-ipstore.}}
An IAC leak is reported by IccTA between app \texttt{com.bi.mutabaah.id} and app \texttt{jp.benishouga.clipstore}.
The \emph{source} method \texttt{findViewById} is called in component \texttt{com.bi.mutabaah.id.activity.Statistic}, where the data of a \texttt{TextView} is obtained.
Then the data is stored into an intent with two extras named \texttt{android.intent.extra.SUBJECT} and \texttt{android.intent.extra.TEXT}.
After that, \texttt{startActivity} is used to send the data to app \texttt{jp.benishouga.clipsto\-re}, which extracts the data from the intent with the same extra names and writes all the data into a file named \texttt{clip.txt} under path \texttt{/data/data/jp.benishouga.clipstore/files}.

\textbf{Conclusion.}
IccTA finds leaks in real-world apps in a reasonable amount of time.
Nevertheless, IccTA only detects a single IAC leak.
This is an indication that inter-application leaks are rare.

\subsection{RQ3: Compare with Other academic Tools}
\label{subsec:comparing}

We identify two academic tools able to deal with ICC leaks: 
SCanDroid~\cite{scandroid} and SEFA~\cite{wu2013impact}.
However, ScanDroid fails to report any leaks and SEFA is not available. 
As a result, we were not able to evaluate them on DroidBench. 

To answer the research question, we focus and discuss some key aspects of the various approaches. 
SCanDroid and SEFA both use a \textit{path matching} approach,
which computes paths for all individual components and then combines some paths together, the decision of combining two paths or not is given by a matching algorithm. 
A \textit{path matching} approach presents at least two main drawbacks.

First, even if the taint analysis is done for each component, the context of the analysis is lost when SCanDroid and SEFA combine the taint paths, since the analysis is performed before the combination of the paths. 
IccTA does not present this problem because it connects the components at the code level and then performs the analysis. Thus, it keeps the data-flow between two components.
Losing the context decreases the precision of the tool. 
Indeed, an \texttt{Intent} can carry data, i.e., it may contain a lot of extras key/value pairs but only part of them are sensitive. A precise tool needs to distinguish them to avoid false positive.
For a path matching approach, it is not easy to distinguish them because they do not keep the state of \texttt{Intent} when matching two available paths.

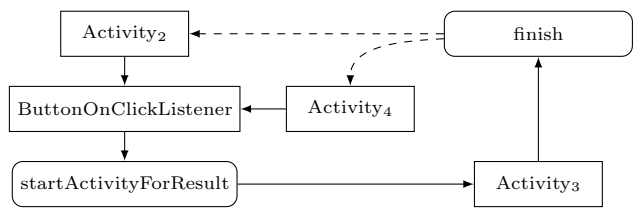
\begin{figure}
\begin{center}
\begin{tikzpicture}
 \tikzstyle{nodeS} = [draw, minimum width=2.5cm, minimum height=.6cm, font=\scriptsize]
 \tikzstyle{nbr} = [solid, midway, above, circle, draw, inner sep=.12mm, yshift=2mm, font=\scriptsize];
 \tikzstyle{emptyS} = [inner sep=0cm, outer sep=0cm]

\path
(0, 0)node[nodeS, minimum width=1.7cm] (a2) {Activity$_2$}
++(0, -1cm) node[nodeS] (onClick) {ButtonOnClickListener}
++(0, -1cm) node[nodeS, rounded corners] (start) {startActivityForResult}

++(3cm, 1cm) node[nodeS, minimum width=1.7cm] (a4) {Activity$_4$}

++(2.5cm, 1cm) node[nodeS, rounded corners] (finish) {finish}
++(0, -2cm) node[nodeS, minimum width=1.7cm] (a3) {Activity$_3$}
;

\draw[-latex] (a2) -- (onClick);
\draw[-latex] (a4) -- (onClick);
\draw[-latex] (onClick) -- (start);
\draw[-latex] (start) -- (a3);
\draw[-latex] (a3) -- (finish);
\draw[-latex, dashed] (finish) -- (a2);
\draw[-latex, dashed] (finish) to[out=183, in=90] (a4);
\end{tikzpicture}
\end{center}

\caption{The problem of using path matching approach for \texttt{startActivityForResult}}
\label{fig:startActivityForResult_problem}
\end{figure}

Second, some specific ICC methods such as \texttt{startActivityForResult} are difficult to handle with a matching algorithm. 
It will become even worse when the special ICC methods exist in a class which is invoked by multiple components.
Suppose a component \texttt{Activity$_4$} also uses the class \texttt{ButtonOnClickListener} shown in Listing~\ref{motivatingExampleCode} to communicate with other components.
We present this scenario in Figure~\ref{fig:startActivityForResult_problem}.
A path matching approach first finds a path from \texttt{startActivityForResult} to \texttt{Activity$_3$}.
After the \texttt{finish} method of \texttt{Activity$_3$} is called, the \texttt{onActivityResult} method of the source component is invoked by the Android system.
The problem is that it is difficult to know which component (\texttt{Activity$_2$} or \texttt{Activity$_4$}) is the source because they both use the same class \texttt{ButtonOnClickListener} where the Intent is created. 
In fact, It is very difficult to statically resolve this problem since it is caused by the mechanism of dynamic binding of Android (or Java).
In our approach, IccTA resolves this problem by explicitly calling the appropriate \texttt{onActivityResult} method (see Figures~\ref{fig:mechanism_startActivityForResult} and \ref{fig:instrumenting_startActivityForResult}) of the source component (\texttt{Activity$_2$} or \texttt{Activity$_4$}) thanks to the helper class \texttt{IpcSC}.

\textbf{Conclusion.}
Even if we were not able to evaluate state-of-the-art tools detecting ICC leaks (SCanDroid and SEFA), IccTA seems to be more precise mainly because it keeps the context between components unlike \textit{path matching} approaches. 

 \section{Limitations}
\label{limitations_future_work}

In this section, we discuss the limitations of IccTA.

\textbf{FlowDroid.} IccTA is based on FlowDroid to perform static taint analysis and thereby shares the same limitations of FlowDroid.
IccTA resolves reflective calls only if their arguments are string constants.
It is also oblivious to multi-threading.
We experienced that FlowDroid cannot properly analyze some apps (too much memory consumption or hangs).
We start by analyzing a set of 5000 and keep only 3000 apps that work with FlowDroid.
Running IccTA on a big server could significantly decrease the number of falling analysis. 
Moreover, we are very confident that the next release of FlowDroid will resolve this problem.

\textbf{Epicc.} IccTA relies on Epicc to compute links between components.
Since Epicc does not handle URIs, it fails to find ICC links for \texttt{ContentProvider} and yields false positives for the other three types of components when they communicate using URIs.
In practice the number of links is huge due to the false positives.
We check the links (intents and intent filters) and only keep the ones not using URIs.

\textbf{IccTA.} 
At the moment IccTA does not handle some rarely used ICC methods such as \texttt{sendActivities} or \texttt{sendOrderedBroadcastAsUser}.
Data send between component with an intent, is represented as key/value pairs.
When a tainted data is put in the intent, IccTA taints all key/value pairs.
This could result in false positives if a tainted data is put in an intent and, in the receiving component, a non-tainted data is retrieved from the intent and flows to a sink.

\textbf{Native Code.} Some Android application are packaged with native code.
IccTA only analyzes the dex file containing the Dalvik bytecode.
 
 \vspace{-3mm}
 \section{Related Work}
\label{relatedwork}

As far as we know, \tool{} is the first approach to seamlessly connect Android components through code instrumentation in order to perform ICC based static taint analysis.
By using a code instrumentation technique, the state of the context and data (e.g. an \textit{Intent})  is transferred between components.
To the best of our knowledge, there is no other existing static approach to detect Android privacy leaks tackling the ICC problem and keeping state between components.

\textbf{Static Analyses.}
There are several approaches using static analysis to detect privacy leaks.
PiOS~\cite{egele2011pios} uses program slicing and reachability analysis to detect the possible privacy leaks.
TAJ~\cite{tripp2009taj} uses the same taint analysis technique to identify privacy leaks in web applications.
However, these approaches introduce a lot of false positives.
CHEX~\cite{chex} is a tool to detect component hijacking vulnerabilities in Android applications by tracking taints between sensitive sources and externally accessible interfaces.
However, it is limited to at most 1-object-sensitivity which leads to imprecision in practice.
LeakMiner and AndroidLeaks state the ability to handle the Android Lifecycle including callback methods, but the two tools are not context-sensitive which precludes the precise analysis of many practical scenarios.
FlowDroid~\cite{steven2014pldi} introduces a highly precise taint analysis approach with low false positive rate, but it does not identify ICC based privacy leaks.
\tool{} performs an ICC based static taint analysis by instrumenting the code of the original app while keeping the precision high.

ComDroid~\cite{comdroid} and Epicc~\cite{octeau2013effective} are two tools that tackle ICC problem, but they mainly focus on ICC vulnerabilities and do not taint data.

SCanDroid~\cite{scandroid} is a tool for analyzing ICC based privacy leaks.
It prunes all call edges to Android OS methods and conservatively assumes the base object, the parameters and the return value to inherit taints from arguments.
This approach is much less precise than our tool since we model all the Android OS methods (except native methods) with our dummy main method in the control-flow graph.
Another tool SEFA~\cite{wu2013impact} also resolves the ICC problem. It performs a system-wide data-flow analysis to detect possible vulnerabilities (e.g., passive content leaks).
Both SCanDroid and SEFA use a matching approach to analyze inter-component leaks.
SCanDroid defines all the methods importing data to an app as \emph{inflow} methods and all the methods exporting data from an app as \emph{outflow} methods.
Then, it matches the \emph{inflow} and the \emph{outflow} methods to connect two components.
SEFA defines ICC methods as \emph{bridge-sinks} to distinguish with the \emph{sensitive-sinks}.
It uses the \emph{bridge-sinks} to match with other components and thereby connecting two components.
As we mentioned before, the matching approach has some drawbacks compared to our instrumenting approach.
Therefore, even if we were not able to evaluate SCanDroid and SEFA on DroidBench, it comes that IccTA is more precise by design. 

AsDroid~\cite{asdroid} and AppIntent~\cite{yang2013appintent} are another two tools using static analysis to detect privacy leaks in Android apps.
Both of them try to analyze the intention of privacy leaks. 
Analyzing the leaking intention is out of scope of this paper.
However, we think it is necessary to distinguish whether a privacy leak is intended or not.
We take this as our further work.

\textbf{Dynamic Analyses.}
Dynamic taint analyses techniques, on the other hand, track sensitive data at runtime.
TaintDroid~\cite{enck2010taintdroid} is one of the most sophisticated dynamic taint tracking systems.
It tracks flows of private data of third-party apps.
CopperDroid~\cite{reina2013system} is another dynamic testing tool which observes interactions between the Android components and the Linux system to reconstruct high-level behavior and uses some special stimulation techniques to exercise the app to find malicious activities.
Several other systems, including AppFence~\cite{hornyack2011these}, Aurasium~\cite{Aurasium}, AppGuard~\cite{appguard} and BetterPermission~\cite{bartel2012invivo} try to mitigate the privacy leak problem by dynamically monitoring the tested apps.

However, those dynamic approaches can be fooled by special designed methods to circumvent security tracking~\cite{sarwar2013effectiveness}.
Thus, dynamic tracking approaches may miss some data leaks and yield an under-approximation.
On the other hand, static analysis approaches may yield an over-approximation because all the application's code is analyzed even code that will never be executed at runtime.
Both approaches are complementary when analyzing Android applications for data leaks.

 \vspace{-3mm}
 \section{Conclusion}
\label{conclusion}

This paper addresses the major challenge of performing data-flow analysis across multiple components or multiple applications.
We have presented IccTA\footnote{Our experimental results and IccTA itself are available at https://sites.google.com/site/icctawebpage.}, an ICC based taint analysis tool able to perform such analysis. In particular, we demonstrate that IccTA can detect ICC based privacy leaks by providing a highly precise control-flow graph through instrumentation of the code of applications.
Unlike previous approaches, IccTA enables a data-flow analysis between two components and adequately models the lifecycle and callback methods to detect ICC based privacy leaks.
When running IccTA on DroidBench, it reaches a precision of \precisionDroidbench{}.
When running IccTA on three thousands applications randomly selected from the Google Play store as well other third-party markets, it detects 130 inter-component based privacy leaks in 12 applications.
Other existing privacy detecting tools (e.g., AndroidLeaks) could benefit by implementing our approach to perform ICC and IAC based privacy leaks detection.

\bibliographystyle{abbrv}
\bibliography{leaks} 
\balance

\end{document}